\documentclass[preprint,aps,prd,superscriptaddress,nofootinbib,tightenlines]{revtex4}

\usepackage{amsfonts}
\usepackage{multirow}
\usepackage{mathrsfs}
\usepackage{graphicx}
\usepackage{amsmath}
\usepackage{amssymb}
\usepackage{bm}
\usepackage{bbm}
\usepackage{color}
\usepackage{xcolor}
\usepackage{float}
\usepackage{slashed}
\usepackage{ulem}

\def\XXint#1#2#3{{\setbox0=\hbox{$#1{#2#3}{\int}$ }
\vcenter{\hbox{$#2#3$ }}\kern-.6\wd0}}

\newcommand{\nn}{\nonumber}

\newcommand{\beq}{\begin{equation}}
\newcommand{\eeq}{\end{equation}}
\newcommand{\bqa}{\begin{eqnarray}}
\newcommand{\eqa}{\end{eqnarray}}
\newcommand{\bseq}{\begin{subequations}}
\newcommand{\eseq}{\end{subequations}}

\begin{document}
\preprint{JLAB-THY-24-3998}
\title{\mbox{}\\[10pt]
Hard-scattering approach to strongly hindered electric dipole transitions between heavy quarkonia}

\author{Cai-Ping Jia\footnote{jiacp20@lzu.edu.cn}}
\affiliation{School of Nuclear Science and Technology, Lanzhou University, Lanzhou 730000, China\vspace{0.2cm}}
\affiliation{Institute of High Energy Physics, Chinese Academy of Sciences, Beijing 100049, China\vspace{0.2cm}}

\author{Yu Jia\footnote{jiay@ihep.ac.cn}}
\affiliation{Institute of High Energy Physics, Chinese Academy of
Sciences, Beijing 100049, China\vspace{0.2cm}}
\affiliation{School of Physical Sciences, University of Chinese Academy of Sciences, Beijing 100049, China\vspace{0.2cm}}

\author{Junliang Lu\footnote{lujl@ihep.ac.cn}}
\affiliation{Institute of Particle and Nuclear Physics, Henan Normal University, Xinxiang 453007, China\vspace{0.2cm}}
\affiliation{Institute of High Energy Physics, Chinese Academy of Sciences, Beijing 100049, China\vspace{0.2cm}}

\author{Zhewen Mo\footnote{mozw@itp.ac.cn}}
\affiliation{CAS Key Laboratory of Theoretical Physics, Institute of Theoretical Physics,
	Chinese Academy of Sciences, Beijing 100190, China\vspace{0.2cm}}
\affiliation{Institute of High Energy Physics, Chinese Academy of
	Sciences, Beijing 100049, China\vspace{0.2cm}}

\author{Jia-Yue Zhang\footnote{jzhang@jlab.org}}
\affiliation{Theory Center, Jefferson Lab, Newport News, Virginia 23606, USA\vspace{0.2cm}}
\affiliation{Institute of High Energy Physics, Chinese Academy of
	Sciences, Beijing 100049, China\vspace{0.2cm}}

\date{\today}

\begin{abstract}
The conventional wisdom in dealing with electromagnetic transition between heavy quarkonia is the multipole expansion,
when the emitted photon has a typical energy of order quarkonium binding energy. Nevertheless, in the case when the energy carried by
the photon is of order typical heavy quark momentum,
the multipole expansion doctrine is expected to break down.
In this work, we apply the ``hard-scattering'' approach originally developed to tackle the strongly hindered magnetic dipole ($M1$) transition
[Y.~Jia {\it et al.}, Phys. \ Rev. \ D. 82, 014008 (2010)] to the strongly hindered electric dipole ($E1$) transition between heavy quarkonia.
We derive the factorization formula for the strongly hindered $E1$ transition rates at the lowest order in velocity and $\alpha_s$
in the context of the non-relativistic QCD (NRQCD), and conduct a detailed numerical comparison with the standard predictions for various bottomonia and
charmonia $E1$ transition processes.
\end{abstract}


 \maketitle

\section{Introduction}

Electromagnetic (EM) transitions provide important means to unravel the internal structure of quarkonium and search for the new quarkonium states~\cite{QuarkoniumWorkingGroup:2004kpm, Eichten:2007qx}. From the experimental perspective, many EM transition channels of charmonia and bottomonia
have been precisely measured in \texttt{BES}, $B$ factories, and \texttt{CLEO} experiments~\cite{Workman:2022ynf}.
From the theoretical standpoint, the standard approach to tackle the EM transition between quarkonium
is based on the multipole expansion, essentially identical to the well-known methodology in dealing with the atomic EM transition.
The dominant transition mode is the electric dipole ($E1$) transition, where
the orbital angular momentum of heavy quarks is changed by one unit while the spin is conserved.
The subleading transition mode is the magnetic dipole ($M1$) transition, where the orbital angular momentum does not change but the spin
is flipped.

In the standard picture of multipole expansion, the radiated photon is assumed to carry a very small ({\it ultrasoft}) momentum, $k \sim mv^2$,
where $m$ denotes the heavy quark mass, and $v$ signifies the typical velocity of the quark within the quarkonium.
Consequently, the wavelength of the emitted photon is much longer than the size of the quarkonium, so that the transition amplitude can be expressed as
an overlap integral between the wave functions of the initial-state and final-state quarkonia.
The leading $E1$ transition formula has long been known in the spontaneous radiation of atoms about one century ago, as one of the triumphs of the quantum
theory of electromagnetic fields~\cite{Dirac:1927dy}. The corresponding $E1$ transition formula for quarkonium is essentially the same.
The modern treatment of EM transitions of quarkonia is expedited by invoking the machinery of nonrelativistic effective field theory (EFT) approach,
the so-called potential NRQCD (pNRQCD), where the relativistic corrections can be systematically accounted~\cite{Brambilla:2005zw,Brambilla:2012be}.

Since the binding force for an atom in QED is Coulombic,  it is appropriate to assume the energy difference between any two atomic levels is of order $mv^2$.
However, the story is quite different for heavy quarkonium. The interquark potential is Coulombic only at short distance and becomes linearly rising confinement potential
at long distance. As a consequence, the mass difference between the highly-excited quarkonium and the lowest-lying state could be rather large, actually could be arbitrarily
large in the $N_c\to\infty$ limit or if neglecting the string breaking effect.
For example, the mass difference between $\Upsilon(3S)$ and $\eta_b$ is as large as 1 GeV.
In such a case, it is obviously inappropriate to regard the photon as ultrasoft, and the
standard multipole expansion doctrine simply breaks down.

In fact, for the $M1$ transition process $\Upsilon(3S)\to\eta_b+\gamma$, it is more appropriate to regard the photon to be {\it semi-hard} ($k\sim mv$)
rather than ultrasoft. Assuming that perturbative QCD is still applicable at the scale $mv\sim 1$ GeV,
the ``hard-scattering" approach has been developed to tackle a class of strongly hindered
$M1$ transitions exemplified by $\Upsilon(3S)\to\eta_b+\gamma$ in 2009~\cite{Jia:2009yg}. The key idea is that a semi-hard gluon has to be exchanged
between heavy quark and antiquark in such a process, so that the transition amplitude can be expressed as the convolution between the
radial wave functions of the parent and daughter quarkonia and the perturbatively calculable hard-scattering kernel.
Without adjustable parameters, the predictions made in \cite{Jia:2009yg} are found in
decent agreement with the measurements.

The goal of this paper is to extend the theoretical framework in \cite{Jia:2009yg} to the case of the strongly hindered $E1$ transitions
between heavy quarkonia. For concreteness, we concentrate on the $E1$ transitions between $P$-wave and $S$-wave quarkonium states.
We derive the corresponding factorization formula in the context of nonrelativistic QCD (NRQCD),
and infer the ``hard-scattering" kernel at the leading order in $v$ and $\alpha_s$.  We also numerically compare the predictions from this new formalism
with those from the conventional multipole formula.

The rest of the paper is organized as follows.
In Section~\ref{theoretical:setup}, we introduce our theoretical framework to deal with $E1$ transition,
the nonrelativistic EFT with $ SU(3)_c \times\,U(1)_{\rm em}$ gauge group.
In Section~\ref{Conventional:E1:formula}, we recap the standard $E1$ transition formula from the conventional multipole expansion context.
Section~\ref{sec:main:body} is the main body of the paper. Starting from the NREFT,
we apply the hard-scattering mechanism to derive the formula for strongly hindered
$E1$ transition, taking the $P$- to $S$-wave quarkonium transition as an explicit example.
In Section~\ref{sec:numerical:results}, we present the numerical predictions for various $E1$ transition processes in bottomonium and charmonium family,
comparing the predictions from the new mechanism with the conventional one.
We finally summarize in Section~\ref{sec:summary}.
In Appendix we illustrate that identical formula for strongly hindered $E1$ transition can also be derived in the underlying relativistic field theory.

\section{Theoretical setup: nonrelativistic effective field theory }
\label{theoretical:setup}

A heavy quarkonium is a nonrelativistic bound state that contains towers of distinct scales: the heavy quark mass $m$,
its typical momentum $mv$, and its typical kinetic energy $m v^2$, obeying the hierarchy $m\gg mv\gg mv^2$.
Needlessly to say, there also exists the characteristic
QCD scale $\Lambda_{\rm QCD}$. From phenomenological potential models, one empirically accepts that $v^2\approx 0.3$ for charomonium, and $v^2\approx 0.1$ for bottomonium.
The momentum scale $mv\sim 0.9$ GeV for charmonium and $mv\sim 1.2$ GeV for bottomonium. The typical binding energy
$m v^2\sim \Lambda_{\rm QCD}\approx 0.5 $ GeV for both charmonium and bottomonia.

The field-theoretical tool to tackle heavy quarkonium is an effective field theory dubbed nonrelativistic QCD (NRQCD)~\cite{Caswell:1985ui}, which describes slowly-moving heavy quark and heavy
antiquark to interact with the soft gluons. The predictions for a physical observable can be organized in velocity expansion.
In order to describe the electromagnetic transition between heavy quarkonium, it is convenient to enlarge NRQCD by including the low-energy photon as
the explicit degree of freedom.  For this purpose, we promote the NRQCD to a nonrelativistic EFT (NREFT)
with gauge group $ SU(3)_c \times\,U(1)_{\rm em}$.
Since the $E1$ transition does not flip the heavy quark/antiquark spin, suffices it to retain only the lowest-order
NREFT Lagrangian:
\beq
\mathcal{L}_\text{NREFT}  = \psi^\dagger \left( iD_0+\frac{\boldsymbol{D}^2}{2m} \right)\psi
+ \chi^\dagger \left( iD_0+\frac{\boldsymbol{D}^2}{2m} \right)\chi + {\cal L}_{\rm light}+ \cdots,
\label{NREFT:lagrangian}
\eeq
where $\psi$ ($\chi^\dagger$) denotes the Pauli spinor fields that annihilate a heavy quark (antiquark), $m$ signifies the heavy quark mass,
${\cal L}_{\rm light}+$ refers to the standard QCD Lagrangian for gluons and light quarks.
The covariant derivatives $iD_0=i\partial_0-g t^a A_0^a  - e e_q A^{\rm em}_0$, $i\boldsymbol{D}=i\nabla+g T^a \boldsymbol{A}^a + e e_Q \boldsymbol{A}^{\rm em}$,
with $t^a$ ($a=1,\cdots,8$) denoting the $SU(3)_c$ generators in fundamental representation, and $e e_Q$ denoting the electric charge of the heavy quark.
For simplicity, we have neglected all the higher-order terms in heavy quark velocity expansion.
Note the NREFT at lowest order in $v$ emerges a rigorous heavy quark spin symmetry, which is manifest in the fact
that \eqref{NREFT:lagrangian} does not explicitly entail Pauli matrices.

One obtains the NRQCD Lagrangian by integrating out the relativistic quantum fluctuation (greater or equal to $m$) from QCD.
As elucidated by Beneke and Smirnov~\cite{Beneke:1997zp},
three distinct dynamical low-energy modes are present in NRQCD,
with the dispersive scaling behaviors specified by~\footnote{Note that the semi-hard mode is often dubbed the soft mode in the literature.}
\bseq
\begin{align}
    \text{semi-hard mode}:& \quad k^0\sim|\boldsymbol{k}|\sim m v;\\
    \text{potential mode}:& \quad k^0\sim mv^2,\qquad |\boldsymbol{k}|\sim m v;\\
    \text{ultrasoft mode}:& \quad k^0\sim|\boldsymbol{k}|\sim m v^2.
\end{align}
\eseq

\section{Review of conventional $E1$ transition formula}
\label{Conventional:E1:formula}

It is the $e e_Q \boldsymbol{A}^{\rm em}\cdot \boldsymbol{\nabla}$ term in the NREFT Lagrangian \eqref{NREFT:lagrangian} that mediates the $E1$ transition.
In the standard textbook treatment, the emitted photon is assumed to be ultrasoft ($|\boldsymbol{k}|\sim m v^2$),
whose wavelength is much longer than the typical quarkonium size ($\sim 1/mv$). Thus it is legitimate to carry out the multipole expansion and the
leading EM transition is dominated by the $E1$ transition. In such case, the semi-hard quanta are irrelevant and can be integrated out,
so that one descends from NRQCD to potential NRQCD (pNRQCD),
which turns out to be the appropriate EFT to describe the electromagnetic transition between quarkonia.

The spin-averaged quarkonium $E1$ transition rate in the traditional multipole expansion framework reads~\cite{QuarkoniumWorkingGroup:2004kpm}
\beq
    \Gamma({}^{2S+1} L_J  \to {}^{2S^\prime+1} L^\prime_{J^\prime} + \gamma)
    =  \frac{4\alpha e_Q^2}{3}  \left(2J^\prime+1\right)
    S_{if}^E k^3
    \left|\varepsilon_{if}\right|^2,
\label{conventional:E1:width formula}
\eeq
where a quarkonium state is characterized by the principle quantum number $n$, the total angular momentum $J$,
the orbital angular momentum $L$ and total spin $S$. The statistical factor $S_{if}^E=S_{fi}^E$ is
\begin{equation}
    S_{if}^E
    =
    \text{max}\left(L,L^\prime\right)
    \begin{Bmatrix}
    J & 1 & J^\prime\\
    L^\prime & S & L\\
   \end{Bmatrix}^2,
\end{equation}
where the last term in the brackets denotes the 6-$j$ symbol. The QCD dynamics is encapsulated in the overlap integral $\varepsilon_{if}$,
which at the lowest order in $v$ is given by
\begin{equation}
\varepsilon_{if} = \int_0^\infty\!\! dr \, r^3\, R_{nL}(r) R_{n^\prime L^\prime}(r),
\label{conventional:E1:overlap:integral}
\end{equation}
where $R_{nL}(r)$ denotes the radial wave function with principle quantum number $n$ and orbital angular momentum $L$.
For the ${\cal O}(v^2)$ corrections to the quarkonium $E1$ transition processes in the framework of pNRQCD, we refer the
interested readers to \cite{Brambilla:2012be}.

\section{strongly hindered $E1$ transition between $P$-wave and $S$-wave quarkonia}
\label{sec:main:body}

In contrast to the atomic systems, the mass difference between highly excited quarkonium and the lowest-lying quarkonium state could be as large as 1 GeV (imagine the mass difference
between $\chi_{bJ}(3P)$ and $\Upsilon(1S)$).
In such a situation, it is much more appropriate to regard the emitted photon as {\it semi-hard} rather than ultrasoft,
and one definitely ought to give up the traditional multipole expansion approach.
Upon emitting a semi-hard photon, an almost on-shell (potential) heavy quark will inevitably become semi-hard,
which bears a virtuality of order-$(m v)^2$ and can only survive in a relatively short time ($~1/mv$) according to the uncertainty principle.
In order to bring this virtual heavy quark back to almost on-shell (potential), one semi-hard gluon must be exchanged between it and the spectator heavy anti-quark.
This is the essential physical consideration that leads to the ``hard-scattering" mechanism~\cite{Jia:2009yg}.
This new mechanism has been first applied to predict the strongly hindered $M1$ transition process $\Upsilon(3S)\to \eta_b\gamma$,
and found notable phenomenological success. It is the major goal of this work to apply this ``hard-scattering" mechanism to tackle the
strongly hindered $E1$ transition between heavy quarkonia.

\subsection{Spin-conserving quark amplitude from ``hard-scattering" mechanism}
\label{amplitude:quark-level:NREFT}

\begin{figure*}[htbp]
\begin{center}
\includegraphics[clip,width=1\textwidth]{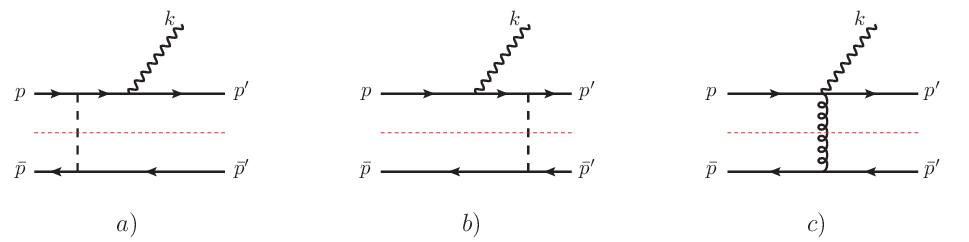}
\caption{Three lowest-order NREFT diagrams that contribute to the strongly hindered $E1$ transition between heavy quarkonia arising from the hard-scattering mechanism. The vertical dashed and curly lines
signify the temporal and spatial gluon propagators. The horizontal thin dashed cut emphasizes that
the upper half of the diagrams are reminiscent of the Compton scattering process,
while the lower half of the diagrams are simply a quark-gluon vertex.
For simplicity, we have omitted three mirror diagrams where the photon is emitted from the heavy anti-quark line. }
\label{Hard-scattering:E1:NREFT:diagram}
\end{center}
\end{figure*}

Let us first consider the on-shell amplitude for $Q(p, s_1)\overline{Q}(\bar{p},s_2)\to Q(p^\prime,s_1^\prime)\overline{Q}(\bar{p}^\prime,s_2^\prime)+\gamma(k)$, where
all the incoming and outgoing quarks and antiquarks move nonrelativistically.
For simplicity, we choose to work in the rest frame of the parent quarkonium, and assume the photon moves along the $\hat{\bf z}$ axis.
$s_1$, $s_2$ refer to the magnetic quantum number of the incoming $Q$ and $\overline{Q}$, and
$s^\prime_1$, $s^\prime_2$ refer to the magnetic quantum number of the outgoing $Q$ and $\overline{Q}$.
	For our interest, we are only concerned in the case where the final-state $Q\overline{Q}$ pair inherits the same total spin from the initial-state
$Q\overline{Q}$ pair.
Denoting the total and relative momenta of the incoming (outgoing) heavy quark-antiquark pair by $P$ and $q$ ($P^\prime$ and $q^\prime$),
respectively, we then have
\bseq
\bqa
& & p=\frac{P}{2}+q,\qquad \bar{p}=\frac{P}{2}-q;
\\
 && p' =\frac{P^{\prime}}{2}+q^{\prime},\qquad \bar{p}^\prime=\frac{P^{\prime}}{2}-q^{\prime}.
\eqa
\label{quark:antiquark:momentum:kinematics}
\eseq
The on-shell condition demands that $P\cdot q=0$ and $P^\prime\cdot q^\prime=0$.
For the incoming quark-antiquark pair, we have $P^\mu = (M_i,{\bf 0})$ and $q^\mu=(0, {\bf q})$,
with $M_i\approx 2 m$ signifying the mass of the parent quarkonium. It is natural to assume ${\bf q}\sim {\cal O}(mv)$.
We will regard the emitted photon as semi-hard, {\it i.e.}, ${\bf k}\sim {\cal O}(mv)$.
Since the daughter quarkonium recoils against the emitted photon, the daughter quarkonium acquires a nonzero yet small
spatial momentum ${\bf P}^\prime = - {\bf k}\sim {\cal O}(mv)$. As a consequence of $P^\prime\cdot q^\prime=0$,
the relative momentum for the outgoing quark-antiquark pair has a typical
spatial component ${\bf q}^\prime \sim {\cal O} (mv)$, yet acquires a strongly suppressed temporal component
$q^{\prime 0} \sim {\cal O} (mv^2)$.

The punch line of the ``hard-scattering" mechanism is that a semihard gluon must be
exchanged between the heavy quark and heavy antiquark.
Half of the lowest-order Feynman diagrams have been depicted in Fig.~\ref{Hard-scattering:E1:NREFT:diagram}.
The Feynman rules can be straightforwardly read off from the NREFT Lagrangian \eqref{NREFT:lagrangian}.
Although it is a popular practice to use Coulomb gauge in computing the NRQCD diagrams,
it turns out that Feynman gauge is a more convenient choice to handle the process entailing
a semi-hard gluon exchange. The corresponding NREFT amplitudes affiliated with three diagrams in
Fig.~\ref{Hard-scattering:E1:NREFT:diagram} are
\bseq
\bqa
\label{eq:amp:a}
 &&   i \mathcal{A}_a
 = \xi^{s'_1\dagger} \left( ie e_Q { 2 {\bf p}^\prime \cdot \boldsymbol{\varepsilon}_\gamma^* \over 2m}\right)
    {i\over l_{Q1}^0-\frac{\boldsymbol{l}_{Q1}^2}{2m}}
    (ig_s t^a) \xi^{s_1} \left({-i\over l_g^2+ i \epsilon} \right) \eta^{s_2^{\dagger}} (ig_s t^a) \eta^{s'_2},
\\
&& i \mathcal{A}_b =  \xi^{s'_1\dagger}
     (ig_s t^a)
     \frac{i}{l_{Q2}^0-\frac{\boldsymbol{l}_{Q2}^2}{2m}}
    \left( ie e_Q { 2 {\bf p} \cdot \boldsymbol{\varepsilon}_\gamma^* \over 2m}\right)
 \xi^{s_1} \left({-i\over l_g^2+i\epsilon} \right) \eta^{s_2^{\dagger}}  (ig_s t^a) \eta^{s'_2},
\\
&& i \mathcal {A}_{c} =
  \xi^{s'_1\dagger} \,\left( {ig_{s} e e_Q \over 2m} t^a \right)\, \xi^{s_1}
  \left( {i \over l_g^{2}+i\epsilon} \right)
    \eta^{s_2^{\dagger}} \,\left(
    ig_{s}{(\bar{\bf p}+\bar{\bf p}^\prime)\cdot \boldsymbol{\varepsilon}_\gamma^* \over 2m} t^{a}\right) \eta^{s'_2},
\eqa
\label{quark-level:amplitude:NREFT}
\eseq
where the transversity condition ${\bf k}\cdot \boldsymbol{\varepsilon}_\gamma^*=0$ has been utilized. For simplicity, we have suppressed the color indices associated with quark Pauli spinors.

The momenta carried by the quark propagators and gluon propagators in Fig.~\ref{Hard-scattering:E1:NREFT:diagram} are
\bseq
\bqa
&&
l^0_{Q1} \equiv p^{\prime 0}+ |{\bf k}|-m \sim {\cal O}( m v),  \qquad  {\bf l}_{Q1}= {\bf p}^\prime + {\bf k}\sim {\cal O}( m v),
\\
&&
l^0_{Q2} \equiv p^0 - |{\bf k}|-m\sim {\cal O}( m v),  \qquad  {\bf l}_{Q2}= {\bf p} - {\bf k}\sim {\cal O}( m v),
\\
&& l_{g}\equiv p-p^\prime -k = q- q^\prime -{k\over 2}\sim {\cal O}( m v).
\eqa
\label{quark:gluon:propagators:momenta}
\eseq
Note we have subtracted the quark mass in the zeroth-component of the quark 4-momentum,
in accordance with the nonrelativistic dispersion relation.
From \eqref{quark:gluon:propagators:momenta} one immediately observes that
the momenta affiliated with both quark and gluon propagators belong to the semi-hard family.
The gluon propagator can be simplified as
\beq
{1\over l_g^2+i\epsilon} = {1\over  (q- q^\prime -{k\over 2})^2+i\epsilon}
\approx
\frac{-1}{({\bf q}'-{\bf q})^2+ {\bf k}\cdot({\bf q}'-{\bf q})-i\boldsymbol{\epsilon}},
\label{expand:soft:gluon:propagator}
\eeq
which scales homogenously as $1/(mv)^2$.

The nonrelativistic quark propagators become inhomogeneous in the soft region. The quark kinetic energy scales as $m v^2$, which is subleading in $v$
compared with $l_{Q1}^0$.
At the lowest order in $v$, it is legitimate to retain only the $l_{Q1}^0$ piece and resulting the following eikonal propagators:
\bseq
\bqa
{1\over l_{Q1}^0-\frac{{\bf l}_{Q1}^2}{2m}} =
{1\over l_{Q1}^0}\left(1+\frac{{\bf l}_{Q1}^2}{2m l_{Q1}^0}+\cdots\right) \approx {1\over l_{Q1}^0} \approx{1\over |{\bf k}|},
\\
{1\over l_{Q2}^0-\frac{{\bf l}_{Q2}^2}{2m}} =
{1\over l_{Q2}^0}\left(1+\frac{{\bf l}_{Q2}^2}{2m l_{Q2}^0}+\cdots\right) \approx {1\over l_{Q2}^0} \approx -{1\over |{\bf k}|},
\eqa
\label{expand:soft:quark:propagator}
\eseq
which simply scales as $1/mv$. This recipe for the soft quark propagator has been expounded in the context of strategy of region~\cite{Beneke:1997jm,Griesshammer:1997wz}.
To put in other word, we can obtain this result by re-interpreting \eqref{NREFT:lagrangian} as the heavy quark effective theory (HQET) Lagrangian~\cite{Eichten:1989zv,Georgi:1990um}
which contains the spin-independent ${\cal O}(1/m)$ operator. The eikonal propagators in \eqref{expand:soft:quark:propagator} are nothing but
the quark propagators given in HQET.
 It is interesting to note that in Fig.~\ref{Hard-scattering:E1:NREFT:diagram}, the upper half diagrams above the
horizontal cut are reminiscent of the soft Compton scattering process, in which the soft massless gauge bosons scatter off
a static heavy quark. In a sense, it is more appropriate to account for such a process from the angle of HQET rather than NRQCD~\footnote{A cautious reader might
worry about the incompatibility between the kinematic setup in
\eqref{quark:antiquark:momentum:kinematics} and the assumption of the semi-hard gluon propagator.
Some inconsistency may arise from the lower half of Fig.~\ref{Hard-scattering:E1:NREFT:diagram}, since it is impossible for
a potential quark to remain potential after being stuck by a semi-hard gluon. The root of this inconsistency is that in order for $k\sim mv$,
the parent quarkonium has to be a highly-excited state, so that at the quark level, the on-shell condition $\bar{p}^2=m^2$ cannot be simultaneously satisfied
in addition to the condition $p^2=m^2$. Nevertheless, one may hypothesize that the spectator antiquark in the initial state is off-shell by an amount of
$(mv)^2$,  or it still remains on-shell albeit with a larger effective mass, $\bar{p}^2\equiv m^2_{\rm eff} \approx m^2+ {\cal O}(mv)^2$. In
any event,  the role of the lowest half diagram of Fig.~\ref{Hard-scattering:E1:NREFT:diagram} is just to generate a simple mass-independent NRQCD amplitude
$\eta^{s_2^{\dagger}} (ig_s t^a) \eta^{s'_2}$. Therefore, we conclude that the conflict between the kinematic assignment in
\eqref{quark:antiquark:momentum:kinematics} and the semi-hard gluon propagator will not spoil the validity of the NREFT amplitude
\eqref{quark-level:amplitude:NREFT}.}.

Inspecting \eqref{quark-level:amplitude:NREFT}, we notice that the amplitude of Fig.~\ref{Hard-scattering:E1:NREFT:diagram}$c)$ (with a seagull vertex)
is suppressed by a factor $\propto |{\bf k}|/m\sim v$ relative to the amplitudes of Fig.~\ref{Hard-scattering:E1:NREFT:diagram}$a)$ and $b)$,
hence can be safely discarded.

One can also employ the equality ${\bf p}^\prime \cdot \boldsymbol{\varepsilon}_\gamma^*= {\bf q}^\prime \cdot \boldsymbol{\varepsilon}_\gamma^*$
and ${\bf p} \cdot \boldsymbol{\varepsilon}_\gamma^*= {\bf q} \cdot \boldsymbol{\varepsilon}_\gamma^*$  in \eqref{quark-level:amplitude:NREFT}, since ${\bf P}^\prime \cdot \boldsymbol{\varepsilon}_\gamma^* = {\bf P} \cdot \boldsymbol{\varepsilon}_\gamma^* =0$.
Further implementing the homogenized  soft gluon and quark propagators given in
\eqref{expand:soft:gluon:propagator} and \eqref{expand:soft:quark:propagator},
we obtain the full quark-level amplitude at the lowest order in $v$:
\beq
{\cal A}[Q\overline{Q}\to Q\overline{Q}+\gamma(k)] =  {e e_Q  g_s^2\over 2m |{\bf k}|} \, \xi^{s'_1\dagger} t^a  \xi^{s_1} \eta^{s_2^{\dagger}} t^a \eta^{s'_2} \,
\boldsymbol{T}({\bf q'}-{\bf q}) \cdot \boldsymbol{\varepsilon}_{\gamma^*},
\label{quark:level:semi:hard:scattering:amplitude}
\eeq
where
\beq
\boldsymbol{T}({\bf w}) =
{  2 {\bf w} \over  {\bf w}^2 + {\bf w}\cdot {\bf k} - i\epsilon}  + ({\bf k}\to -{\bf k}).
\label{kernel:T:definition}
\eeq
The second term originates from those mirror diagrams where the photon is emitted from the heavy antiquark.

It is worth emphasizing that, the above derivation amounts to integrating out the semi-hard quantum fluctuation in perturbative NRQCD.
A tacit assumption is that the perturbation theory is still applicable, {\it i.e.},
the characteristic strong coupling constant $\alpha_s(m v)\ll 1$.
This assumption is expected to be solid for bottomonia, and might be reasonable even for charmonia.

\subsection{Folding hard-scattering kernel with bound-state wave functions}

Having obtained the quark amplitude for $Q\overline{Q}\to Q\overline{Q}+\gamma(k)$,
we proceed to derive the corresponding $E1$ transition rate between two physical quarkonia states.
To form a bound state, the heavy quark and antiquark are assumed to be nearly on-shell (potential),
exchanging infinite ladders of potential gluons.  The corresponding binding dynamics are governed by nonperturbative QCD.
To make a physical prediction for $E1$ transition, we have
to fold the ``hard-scattering" kernel in \eqref{quark:level:semi:hard:scattering:amplitude} with the nonperturbative wave functions
of the initial- and final-state quarkonia.

At the lowest order in $v$, the state vector of a quarkonium $H$ bearing the quantum number $n{}^{2S+1}L_J$,
can be explicitly constructed in terms of the nonrelativistic heavy quark-antiquark basis~\cite{Weinberg:2015cc}
\bqa
&&  \left| H(n^{2S+1}L_J)({\bf P}, J_z) \right \rangle = \sum_{i,j,s_1, s_2, S_z, L_z} {\delta^{ij}\over \sqrt{N_c}}
\left\langle {1\over 2} s_1; {1\over 2}s_2 \bigg\vert S\,S_z \right\rangle
\langle S\,S_z; L\,L_z| J\,J_z \rangle
\nn \\
 & \times & \int\frac{d^3 {\bf q}}{(2\pi)^3}
     \widetilde{R}_{nL}(|{\bf q}|)  Y_{L L_z}(\hat{{\bf q}})
     \left \vert Q^i\left(\frac{\bf P}{2}+ {\bf q}, s_1 \right)\overline{Q}^j \left(\frac{\bf P}{2}-{\bf q},s_2\right)
     \right\rangle,
\label{Construction:quarkonium:state}
\eqa
where $i,j=1,2,3$ denote the color indices, $s_1$ and $s_2$ characterize the magnetic quantum number of the quark and antiquark.
$\widetilde{R}_{nL}(|{\bf q}|)$ denotes the radial wave function for quarkonium in momentum space, $Y_{L L_z}(\hat{\boldsymbol{q}})$
signifies the spherical harmonics.   $ \langle \frac{1}{2} s_{1};\frac{1}{2} s_{2}|S\,S_z\,\rangle$
and $ \langle \,S\,S_z\,;L \,L_z| J\,J_z\,\rangle $ correspond to the Clebsch-Gordan coefficients affiliated with spin-spin coupling and
spin-orbital coupling, where $S_z$ and $L_z$ signify the magnetic quantum numbers
corresponding to the spin and orbital angular momentum.

One readily verifies that the quarkonium state $H$ defined in \eqref{Construction:quarkonium:state}
obeys the standard non-relativistic normalization:
\beq
\langle
H(n^{2S+1}L_{J})\,({\bf P}_1)\, |H(n^{2S+1}L_{J})\,({\bf P}_2)
\rangle = (2\pi)^3 \,\delta^{(3)} ({\bf P}_1- {\bf P}_2),
\eeq
	provided that the momentum-space quarkonium wave function is subject to the following orthogonality condition:
\beq
\int\!\! \frac{d^3 {\bf q}}{(2\pi)^3}  \widetilde{R}_{nL}(|{\bf q}|)  \widetilde{R}^*_{n'L'}(|{\bf q}|) Y_{L L_z}(\hat{{\bf q}}) Y^*_{L' L'_z}(\hat{{\bf q}}) =
\delta_{n n'} \delta_{L L^\prime} \delta_{L_z L_z'}.
\eeq

The selection rule of the $E1$ transition is that the orbital angular momentum changes by one unit while keeping spin unchanged.
For definiteness, in this work we will concentrate on the situation where
the $P$-wave  quarkonium  transitions into a $S$-wave quarkonium.

The $E1$ transition amplitude becomes
\bqa
& & \mathcal{A}[n^{2S+1} P_J\to n'^{2S+1} S_S+\gamma  ] =  \int\frac{d^3 {\bf q}}{(2\pi)^3}\int\frac{d^3 {\bf q'}}{(2\pi)^3}
    \frac{1}{N_c} \sum_{i,s_1, s_2, S_z, L_z}
\left\langle {1\over 2} s_1; {1\over 2}s_2 \bigg\vert S\,S_z \right\rangle
\langle S\,S_z; 1\,L_z| J\,J_z \rangle
\nn\\
   &&  \times \sum_{i',s'_1, s'_2}
\left\langle S' \,S'_z   \bigg\vert  {1\over 2} s'_1; {1\over 2}s'_2 \right\rangle
    \widetilde{R}_{nP}(|{\bf q}|)  Y_{1 L_z}(\hat{{\bf q}})  \widetilde{R}^*_{n'S}(|{\bf q'}|)Y^*_{00}(\hat{{\bf q'}})
\label{E1:amplitude:P:wave:to:S:wave}
\\
&&\times  {\cal A}[Q^i (s_1) \overline{Q}^i(s_2) \to Q^{i'}(s'_1)\overline{Q}^{i'}(s'_2)+\gamma(k)],
\nn
\eqa

The quark and anti-quark can form either spin singlet or spin triplet. It is convenient to adopt the following
spin-spin coupling formula:
\bseq
\bqa
&& \sum_{s_1,s_2} \langle
     \frac{1}{2} s_{1};\frac{1}{2}s_{2}|0\,0
     \rangle
     \xi^{s_{1}}\eta^{s_{2}\dagger}
     = {1\over \sqrt{2}}\,{\bf 1}_{2\times 2},
\qquad
\sum_{s'_1,s'_2}
\langle 0 \,0 \vert  {1\over 2} s'_1; {1\over 2} s'_2 \rangle
\eta^{s'_2}\xi^{s'_1\dagger} =\frac{1}{\sqrt{2}}\,{\bf 1}_{2\times 2},
\\
&& \sum_{s_1,s_2}
     \langle
     \frac{1}{2} s_{1};\frac{1}{2}s_{2}|1\,S_z
    \rangle
    \xi^{s_{1}}\eta^{s_{2}\dagger}
    =
    \frac{1}{\sqrt{2}}
    \boldsymbol{\sigma} \cdot \boldsymbol{\varepsilon}(S_z),
\qquad
\sum_{s'_1,s'_2}
     \langle 1 \,S'_z \vert  {1\over 2} s'_1; {1\over 2} s'_2 \rangle
     \eta^{s'_2}\xi^{s'_1\dagger}
     =\frac{1}{\sqrt{2}} \boldsymbol{\sigma} \cdot \boldsymbol{\varepsilon}^*(S'_z),
\nn
\\
\eqa
\label{spin:spin:coupling:relation}
\eseq
where $\boldsymbol{\varepsilon}(S_z)$ refers to the polarization vector affiliated with the unit spin.

Implementing \eqref{spin:spin:coupling:relation} into \eqref{E1:amplitude:P:wave:to:S:wave}, we find
\bseq
\bqa
& &  \mathcal{A}[n^{1} P_1(J_z) \to n'^{1} S_0+\gamma  ] =  C_F  {e e_Q  g_s^2\over 2m |{\bf k}|}
\int\!\!\! \frac{d^3 {\bf q}}{(2\pi)^3}\int\!\!\! \frac{d^3 {\bf q'}}{(2\pi)^3}
\left({1\over \sqrt{2}}\right)^2 {\rm Tr}[  {\bf 1}^2_{2\times 2} ]
\label{E1:amplitude:1P1:to:1S0}
\\
&\times&
 \widetilde{R}_{nP}(|{\bf q}|)  Y_{1 J_z}(\hat{{\bf q}})  \widetilde{R}^*_{n'S}(|{\bf q'}|)Y^*_{00}(\hat{{\bf q'}}) \,
\boldsymbol{T}({\bf q'}-{\bf q}) \cdot \boldsymbol{\varepsilon}_{\gamma}^*,
\nn\\
& & \mathcal{A}[n^{3}P_J (J_z) \to n'^{3} S_1(S'_z)+\gamma  ] =
C_F  {e e_Q  g_s^2\over 2m |{\bf k}|} \int\!\!\! \frac{d^3 {\bf q}}{(2\pi)^3}\int\!\!\! \frac{d^3 {\bf q'}}{(2\pi)^3}
    \sum_{S_z, L_z}
\langle 1\,S_z; 1\,L_z| J\,J_z \rangle
\label{E1:amplitude:3PJ:to:3S1}
\\
&\times &
    \widetilde{R}_{nP}(|{\bf q}|)  Y_{1 L_z}(\hat{{\bf q}}) \widetilde{R}^*_{n'S}(|{\bf q'}|)Y^*_{00}(\hat{{\bf q'}})\,
\left({1\over \sqrt{2}}\right)^2 {\rm Tr}[ \boldsymbol{\sigma} \cdot \boldsymbol{\varepsilon}^*(S'_z) \boldsymbol{\sigma} \cdot \boldsymbol{\varepsilon}(S_z) ]
\boldsymbol{T} ({\bf q'}-{\bf q}) \cdot \boldsymbol{\varepsilon}_{\gamma}^*,
\nn
\eqa
\label{Pwave:to:Swave:E1:intermediate:formula}
\eseq
where the hard-scattering kernel $\boldsymbol{T}$ is given in \eqref{kernel:T:definition}.
Note we have conducted the color sum using ${\rm Tr}[t^a t^a] = C_F N_c$, with $C_F={N_c^2-1\over 2 N_c}$.

Since the bound-state wave functions are usually solved in the coordinate space,
we proceed to Fourier transforming \eqref{Pwave:to:Swave:E1:intermediate:formula} using
\beq
 \widetilde{R}_{nL}(|{\bf q}|) Y_{L L_z}(\hat{{\bf q}}) =  \int\!\!\! d^{3}{\bf r} \,
e^{-i{\bf q}\cdot {\bf r}} R_{nL}(r) Y_{L L_z}(\hat{{\bf r}}).
\eeq

Since the hard-scattering kernel $\boldsymbol{T}$ depends on the quark relative momenta ${\bf q}$ and ${\bf q'}$
only through the linear combination ${\bf q'}-{\bf q}$,
one readily conducts three-fold integration and only retains the integration over ${\bf r}$:
\bqa
& & \int\!\!\! \frac{d^{3} {\bf q}}{(2\pi)^{3}}
    \int\!\!\! \frac{d^{3} {\bf q'}}{(2\pi)^{3}} \int\!\!\! d^{3}{\bf r} \int\!\!\! d^{3}{\bf r'}
  \: e^{-i {\bf q}\cdot {\bf r}}\, e^{i {\bf q'}\cdot {\bf r'}}
     \, f({\bf r})\,\boldsymbol{T}({\bf q'}-{\bf q}) g({\bf r'})
\nn\\
 & =&   \int\!\!\! d^{3}{\bf r} \, f({\bf r})  \int\!\!\! \frac{d^{3} {\bf w}}{(2\pi)^{3}}  \, \boldsymbol{T}({\bf w})
 \, \int\!\!\! d^{3}{\bf r'} \,  g({\bf r'}) e^{i {\bf w} \cdot{\bf r'}}
 \int\!\!\! \frac{d^{3} {\bf q}}{(2\pi)^{3}} \:
      e^{-i {\bf q}\cdot({\bf r}-{\bf r')} }
\nn\\
& = & \int\!\!\! d^{3}{\bf r}\, f({\bf r})\,g({\bf r}) \int\!\!\! \frac{d^{3} {\bf w}}{(2\pi)^{3}}  \, e^{i {\bf w} \cdot{\bf r}} \boldsymbol{T}({\bf w})
\nn\\
\nn\\
&=&  \int\!\!\! d^{3}{\bf r}\, f({\bf r})\, g({\bf r})\,
	{1\over 2\pi} e^{\frac{i}{2}kr}\, \left[\hat{\bf r} \, {2 i + k r\over r^2}
         \cos\left(\frac{1}{2}kr\cos\theta\right)
         +i\frac{\hat{\bf k}}{r}\sin\left(\frac{1}{2}kr\cos\theta\right)
         \right]
\label{4:fold:integration:simplified}
\eqa
where we have changed the integration variable from ${\bf q'}$ to ${\bf w}= {\bf q'}-{\bf q}$ in the second line.

It is convenient to employ the following representation of the spherical harmonics:
\beq
Y_{00}(\hat{\bf r}) = \sqrt{{1\over 4\pi}}, \qquad Y_{1L_z}(\hat{\bf r}) = \sqrt{{3\over 4\pi}} \hat{\bf r}\cdot\boldsymbol{\varepsilon}(L_z),
\label{spherical:harmonics}
\eeq
where $\boldsymbol{\varepsilon}(L_z)$ refers to the polarization vector associated with the unit orbital angular momentum.

In light of \eqref{4:fold:integration:simplified} and \eqref{spherical:harmonics}, the $P$-to-$S$-wave one-photon transition amplitudes in
\eqref{Pwave:to:Swave:E1:intermediate:formula} can be reduced to
\bseq
\bqa
& & \mathcal{A}[n^{1} P_1(J_z) \to n'^{1} S_0+\gamma  ] =   \sqrt{{1\over 4\pi}} \sqrt{{3\over 4\pi}} C_F  {e e_Q  g_s^2\over 2m |{\bf k}|}
 {\varepsilon}^i(J_z) {\varepsilon}^{j*}_{\gamma}
\nn
\\
&& \times \int\!\!\! dr \,r^2\,
  {2 i + k r\over r^2} e^{\frac{i}{2}kr} R_{nP}(r)    R^*_{n'S}(r) \,
   F^{ij}(kr),
\label{E1:1P1:to:1S0:coordinate:space}\\
& & \mathcal{A}[n^{3}P_J (J_z) \to n'^{3} S_1(S'_z)+\gamma  ] = {\sqrt{3}\over 4\pi}
C_F  {e e_Q  g_s^2\over 2m |{\bf k}|}   \sum_{S_z, L_z}
\langle 1\,S_z; 1\,L_z| J\,J_z \rangle (\boldsymbol{\varepsilon}^*(S'_z) \cdot \boldsymbol{\varepsilon}(S_z))\,
{\varepsilon}^i(L_z) {\varepsilon}^{j*}_{\gamma}
\nn
\\
& & \times   \int\!\!\! d r  \, r^2 {2 i + k r\over r^2} e^{\frac{i}{2}kr} R_{nP}(r) R^*_{n'S}(r) \,
    F^{ij}(kr),
\label{E1:3PJ:to:3S1:coordinate:space}
\eqa
\label{P:to:Swave:E1:formula:coordinate:space}
\eseq
where the rank-2 symmetric tensor can be readily worked out upon performing the angular integration:
\beq
   F^{ij}(kr) \equiv \int\!\!\!{d\Omega\over 4\pi}  \cos\left(\frac{1}{2}kr\cos\theta\right)
     \hat{\bf r}^i  \hat{\bf r}^j  =  {2\over kr} j_1\left({kr\over 2}\right) \delta^{ij} + \left[j_0\left({kr\over 2}\right)-
     {6\over k r} j_1\left({kr\over 2}\right) \right] \hat{\bf k}^i \hat{\bf k}^j.
\eeq
Due to the transversity condition ${\bf k}\cdot \boldsymbol{\varepsilon}_\gamma^*=0$, the second term in the bracket in \eqref{4:fold:integration:simplified} as well as the $\hat{\bf k}^i \hat{\bf k}^j$ piece in $F^{ij}(r)$
yields a vanishing contribution in \eqref{P:to:Swave:E1:formula:coordinate:space}, thus can be discarded.

For the spin-triplet $P$-wave states, there exists the following spin-orbital coupling formula~\cite{Kuhn:1979bb}:
\bseq
\begin{align}
\label{cg:chib0}
    \sum_{S_z, L_z}
    \langle 1\,S_z; 1\,L_z| 0\,0 \rangle {\varepsilon}^{i}(S_z) {\varepsilon}^{j}(L_z)
    &=  -\frac{1}{\sqrt{3}}\delta^{ij},
\\
\label{cg:chib1}
     \sum_{S_z, L_z}
    \langle 1\,S_z; 1\,L_z| 1\,J_z \rangle
    {\varepsilon}^{i}(S_z) {\varepsilon}^{j}(L_z)
     &=
     -\frac{i}{\sqrt{2}}
     \epsilon_{ijk} {\varepsilon}_{^3P_1}^{k}(J_z),
\\
\label{cg:chib2}
\sum_{S_z, L_z} \langle 1\,S_z; 1\,L_z| 2\,J_z \rangle {\varepsilon}^{i}(S_z) {\varepsilon}^{j}(L_z) &=
    {\varepsilon}_{^3P_2}^{ij}(J_{z}),
\end{align}
\label{spin:orbit:recoupling}
\eseq
where  $\boldsymbol{\varepsilon}_{^3P_1}^{k}(J_z)$ denotes the polarization vector of the ${}^3P_1$ state, and
${\varepsilon}_{^3P_2}^{ij}(J_{z})$ signifies the symmetric and traceless polarization tensor of the ${}^3P_2$ state.

With the aid of the spin-orbital coupling relation \eqref{spin:orbit:recoupling}, the  $P$-to-$S$-wave strongly hindered $E1$ transition amplitudes
in \eqref{P:to:Swave:E1:formula:coordinate:space} finally become
\bseq
\label{eq:P2S}
\begin{align}
\mathcal{A}(n^1P_1\to n^{\prime 1}S_0+\gamma)&=\frac{-\sqrt{3}ee_Qg_s^2C_F}{2m\pi}\varepsilon_{nn^\prime}(r)
\,\boldsymbol{\varepsilon}_{^1P_1} \cdot\boldsymbol{\varepsilon}_\gamma^*,
\\
\mathcal{A}(n^3P_0\to n^{\prime 3}S_1+\gamma)&=\frac{ee_Qg_s^2C_F}{2m\pi}\varepsilon_{nn^\prime}(r)\,
\boldsymbol{\varepsilon}^{*}_{^3 S_1}\cdot
\boldsymbol{\varepsilon}^*_\gamma   ,
\\
\mathcal{A}(n^3P_1\to {n^{\prime 3}S_1}+\gamma)&=\frac{\sqrt{3}iee_Qg_s^2C_F}{2\sqrt{2}m\pi}
\varepsilon_{nn^\prime}(r) \, \boldsymbol{\varepsilon}_{^3P_1} \cdot \boldsymbol{\varepsilon}_{^3S_1}^{*} \times \boldsymbol{\varepsilon}_\gamma^{*},
\\
\mathcal{A}(n^3P_2\to {n^{\prime 3}S_1}+\gamma)&=\frac{-\sqrt{3}ee_Qg_s^2C_F}{2m\pi}\varepsilon_{nn^\prime}(r)\, {\varepsilon}_{^3P_2}^{ij}
\varepsilon_{^3S_1}^{i*} \varepsilon_\gamma^{j*},    .
\end{align}
\eseq
where the overlap integral $\varepsilon_{nn^\prime}$ is defined by
\begin{align}
\varepsilon_{nn^\prime}(r) & = \int\!\!
    dr\, r^2 R_{nP}(r) R_{n^\prime S}(r) e^{\frac{i}{2}kr}  \frac{2-ikr}{k^2\,r^3} \, j_1(\frac{kr}{2}).
\label{New:E1:overlap:integral}
\end{align}

Curiously, the overlap integral for the strongly hindered $E1$ transition looks similar to, but somewhat simpler than its
counterpart in the strongly hindered $M1$ transition formula~\cite{Jia:2009yg}:
\beq
\varepsilon^{M1}_{nn^\prime}(r) = \int\!\!
	dr\, r^2 R_{nP}(r) R_{n^\prime S}(r) e^{\frac{i}{2}kr}\frac{1}{r} \left[j_0\left(\frac{kr}{2}\right)-\frac{2-ikr}{kr} \, j_1\left(\frac{kr}{2}\right)\right].
\eeq

Both of the overlap integrals in the strongly hindered $E1$ and $M1$ transitions are complex-valued.
It looks a little bit unusual since we have only considered tree-level diagrams.
From the calculational angle, the occurrence of the imaginary part can be attributed to the fact that the exchanged
semi-hard gluon can become on-shell as ${\bf q}-{\bf q}^\prime = {\bf k}$.
Nevertheless, numerical study reveals that the real part dominates over the imaginary part in the overlap integral for both $E1$ and $M1$ transition processes.
In most quarkonium $E1$ transition processes investigated in Sec.~\ref{sec:numerical:results}, we find
the real part of the overlap integral is several orders of magnitude greater than the imaginary part. Therefore, one may safely discard the contribution of
the imaginary part in the overlap integrals in phenomenological analysis.

It is also interesting to examine the asymptotic expression of the overlap integral in the limit $kr\ll 1$.
For the strongly hindered $M1$ transition, one finds asymptotically	$e^{\frac{i}{2}kr}\frac{1}{r} \left[j_0\left(\frac{kr}{2}\right)-\frac{2-ikr}{kr} \, j_1\left(\frac{kr}{2}\right)\right] \to \quad\frac{2}{3r}+\frac{i k}{2}+\cdots$~\cite{Jia:2009yg}. For the strongly hindered $E1$ transition,
we find asymptotically $e^{\frac{i}{2}kr} \,\frac{2-ikr}{k^2\,r^3} \, j_1(\frac{kr}{2})\to {1\over 3 kr^2} +
{i k^2r\over 72}+\cdots$. At the limit $kr\ll 1$, it appears that the suppression of the imaginary part relative to the real part in the strongly hindered $E1$
transition is much more severe than the strongly hindered $M1$ transition. Finally, substituting this asymptotic expression into \eqref{New:E1:overlap:integral},
the numerical values are very close to the rigorous results of the overlap integrals for various quarkonium $E1$ transition processes.

\subsection{Transition rates for strongly hindered $E1$ processes}

Squaring the amplitudes, summing over final-state spins and averaging upon the initial-state polarizations, and using the formula
\beq
  \Gamma({}^{2S+1} L_J \stackrel{\text{E}_1}{\longrightarrow} {}^{2S+1} L^\prime_{J^\prime} + \gamma)=
  {k\over 2\pi}{1\over 2J+1} \sum_{\rm Pol.}\left|{\cal A}\right|^2,
\eeq
we then obtain
\bseq
\bqa
& & \Gamma(n^3P_{J}\to {n^{\prime 3}S_1}+\gamma)
 = \frac{64e_Q^2\alpha\alpha_s^2C_F^2k}{M_i^2}|\varepsilon_{nn^\prime}|^2,
\\
&& \Gamma(n^3S_1\to n^{\prime 3}P_{J}+\gamma)
     = \frac{64(2J+1)e_Q^2\alpha\alpha_s^2C_F^2k}{3M_i^2}|\varepsilon_{nn^\prime}|^2,
\\
& & 
	\Gamma(n^1P_{1}\to {n^{\prime 1}S_0}+\gamma)
= \frac{64e_Q^2\alpha\alpha_s^2C_F^2k}{M_i^2}|\varepsilon_{nn^\prime}|^2,
\\
&&
	\Gamma(n^1S_0\to n^{\prime 1}P_{1}+\gamma)
= \frac{192e_Q^2\alpha\alpha_s^2C_F^2k}{M_i^2}|\varepsilon_{nn^\prime}|^2,
 \eqa
\label{strongly:hindered:E1:width:formula}
\eseq
for $J=0,1,2$. For simplicity, we have approximated the heavy quark mass $m$ by $M_i/2$, half of mass of the initial-state quarkonium.

Equations~\eqref{strongly:hindered:E1:width:formula} and \eqref{New:E1:overlap:integral} constitute the major new results of this work.

If we neglect the fine splitting among different $P$-wave spin triplet members,
the transition rates satisfy the relation $\Gamma_0:\Gamma_1:\Gamma_2=1:1:1$ for $n^3P_{0,1,2}\to$ $n^{\prime 3}S_1+\gamma$, and
$\Gamma_0:\Gamma_1:\Gamma_2=1:3:5$ for the inverse processes $n^3S_1\to n^{\prime 3}P_{0,1,2}+\gamma$.
These patterns are identical to what are derived from the conventional $E1$ transition formula \eqref{conventional:E1:width formula},
which is a consequence of heavy quark spin symmetry~\cite{Cho:1994ih}.
This is not surprising, since our new $E1$ transition mechanism also respects the heavy quark spin symmetry,
which is a rigorous consequence of the lowest-order NRQCD Lagrangian.

Comparing \eqref{conventional:E1:width formula} and \eqref{strongly:hindered:E1:width:formula}, we immediately notice one important difference between the ``hard-scattering" mechanism and conventional formula, that the sensitivity of the $E1$ transition rate to the photon momentum is much weaker in the former case.

\section{Numerical results}
\label{sec:numerical:results}

In this section, we make a detailed numerical comparison between our new predictions arising from the hard-scattering mechanism
and the conventional ones for $E1$ transition processes between $P$ and $S$-wave bottomonia and charmonia.

The key nonperturbative inputs are the Schr\"{o}dinger wave functions for the parent and daughter quarkonium states.
Assuming that color Coulomb force and linear confinement force are equally important for quarkonium dynamics,
we choose to work with the Cornell potential model~\cite{Eichten:1978tg}:
\begin{equation}
    V_{\text{Cornell}}(r)=\frac{-\kappa}{r}+\frac{r}{a^2},
\end{equation}
where $\kappa=0.52$ and $a=2.34\,\text{GeV}^{-1}$. In solving for the charmonia and bottomonia energy levels and the corresponding wave functions, we also
take $m_c=1.84\;\text{GeV}$ and $m_b= 5.18\;\text{GeV}$~\cite{Eichten:1995ch}. The QCD coupling constant entering
\eqref{strongly:hindered:E1:width:formula} should be evaluated at the characteristic semi-hard scale,
and we take $\alpha_s(m_c v)=0.59$ and  $\alpha_s(m_b v)=0.43$~\cite{Braaten:1997nw}.

\begin{table}[ht]
\caption{The $E1$ transition rates for ${}^3S_1\to {}^3P_J+\gamma$ in bottomonia family, predicted
  from both the ``hard-scattering" mechanism and traditional multipole expansion method.
  We also include experimental measurements whenever available.}
  \label{Table:3S1:3PJ:bottomonia}
  \begin{tabular}{|c|c|c|c|c|} \hline
     process & $k(\text{MeV})$ & $\Gamma_\text{exp}$(keV) & $\Gamma_\text{hard scatt.}$ (keV)& $\Gamma_\text{multipole}$(keV) \\ \hline\hline
    $\Upsilon(3S)\to\chi_{b0}(1P)\gamma$ & 484 & 0.055  &  0.20 & 0.12  \\ \hline
    $\Upsilon(3S)\to\chi_{b1}(1P)\gamma$ & 452 &  0.018 &  0.64 & 0.29  \\ \hline
    $\Upsilon(3S)\to\chi_{b2}(1P)\gamma$ & 434 &  0.203 &  1.11 &  0.43 \\ \hline\hline
    $\Upsilon(3S)\to\chi_{b0}(2P)\gamma$ & 121 &  1.199 &  1.18 & 1.53  \\ \hline
    $\Upsilon(3S)\to\chi_{b1}(2P)\gamma$ & 100 & 2.560  &  4.34 &  2.54 \\ \hline
    $\Upsilon(3S)\to\chi_{b2}(2P)\gamma$ & 86  &  2.662 &  8.42 & 2.69  \\ \hline\hline
    $\Upsilon(2S)\to\chi_{b0}(1P)\gamma$ & 163 & 1.215  &  1.00 & 1.46  \\ \hline
    $\Upsilon(2S)\to\chi_{b1}(1P)\gamma$ & 129 & 2.207  &  3.82 &  2.19 \\ \hline
    $\Upsilon(2S)\to\chi_{b2}(1P)\gamma$ & 110 & 2.287  &  7.48 & 2.28  \\ \hline
  \end{tabular}
\end{table}

\begin{table}[ht]
  \caption{Same as Table~\ref{Table:3S1:3PJ:bottomonia}, but the $E1$ processes are replaced with ${}^3P_J\to {}^3S_1+\gamma$ in bottomonium family.}
  \label{Table:3PJ:3S1:bottomonia}
  \begin{tabular}{|c|c|c|c|c|} \hline
     process & $k(\text{MeV})$ & $\Gamma_\text{exp}$(keV) & $\Gamma_\text{hard scatt.}$ (keV)& $\Gamma_\text{multipole}$(keV) \\ \hline \hline
    $\chi_{b0}(2P)\to\Upsilon(1S)\gamma$ & 744 & -  & 18.05 & 10.74\\ \hline
    $\chi_{b1}(2P)\to\Upsilon(1S)\gamma$ & 764 &  - &  17.54 & 11.64\\ \hline
    $\chi_{b2}(2P)\to\Upsilon(1S)\gamma$ & 777 & -  & 17.23 & 12.25\\ \hline\hline
    $\chi_{b0}(2P)\to\Upsilon(2S)\gamma$ & 208 &  - &  12.44 &  9.22 \\ \hline
    $\chi_{b1}(2P)\to\Upsilon(2S)\gamma$ & 229 & -  &  11.25 & 12.40  \\ \hline
    $\chi_{b2}(2P)\to\Upsilon(2S)\gamma$ & 243 & -  & 10.61 &  14.75 \\ \hline\hline
    $\chi_{b0}(1P)\to\Upsilon(1S)\gamma$ & 391 & -  &  44.82 &  20.00\\ \hline
    $\chi_{b1}(1P)\to\Upsilon(1S)\gamma$ & 424 & -  & 41.24 & 25.43\\ \hline
    $\chi_{b2}(1P)\to\Upsilon(1S)\gamma$ & 442 &  - & 39.48 & 28.85\\ \hline
  \end{tabular}
\end{table}

\begin{table}[ht]
  \caption{Same as Table~\ref{Table:3PJ:3S1:bottomonia}, but the $E1$ processes are replaced with ${}^1P_1\to {}^1S_0+\gamma$ in bottomonium family.}
  \label{Table:1P1:1S0:bottomonia}
  \begin{tabular}{|c|c|c|c|c|}
    \hline
     process & $k(\text{MeV})$ & $\Gamma_\text{exp}$(keV)& $\Gamma_\text{hard scatt.}$(keV)& $\Gamma_\text{multipole}$(keV)\\ \hline\hline
    $h_b(2P)\to\eta_b(1S)\gamma$ & 825 & -  & 16.37  &  14.64 \\ \hline
    $h_b(2P)\to\eta_b(2S)\gamma$ & 258 & -  & 10.04  &  17.58 \\ \hline
    $h_b(1P)\to\eta_b(1S)\gamma$ & 487 &  - & 36.08  &  38.75 \\ \hline
    $\eta_b(2S)\to h_b(1P)\gamma$ & 99 &  - & 15.03  &  3.01 \\ \hline
  \end{tabular}
\end{table}

\begin{table}[ht]
  \caption{The $E1$ transition rates in charmonium family, predicted
  from both the ``hard-scattering" mechanism and traditional multipole expansion method.
  We also include experimental measurements whenever available.}
  \label{Table:E1:charmonia}
  \begin{tabular}{|c|c|c|c|c|} \hline
     process & $k(\text{MeV})$ & $\Gamma_\text{exp}$(keV)& $\Gamma_\text{hard scatt.}$ (keV)& $\Gamma_\text{multipole}$(keV)\\ \hline\hline
    $\chi_{c0}(1P)\to J/\psi\gamma$ & 303 &  151.2 & 436.29  & 141.07  \\ \hline
    $\chi_{c1}(1P)\to J/\psi\gamma$ & 390 & 288.12  &  329.41 & 299.29  \\ \hline
    $\chi_{c2}(1P)\to J/\psi\gamma$ & 429 &  374.30 & 295.12  & 400.66  \\ \hline\hline
    $\psi(2S)\to\chi_{c0}(1P)\gamma$ & 261 & 28.783  &  8.67 &  50.52 \\ \hline
    $\psi(2S)\to\chi_{c1}(1P)\gamma$ & 171 &  28.665 &  41.54 & 42.49  \\ \hline
    $\psi(2S)\to\chi_{c2}(1P)\gamma$ & 128 & 27.989  & 94.11  &  29.58 \\ \hline\hline
    $h_c(1P)\to\eta_c(1S)\gamma$ & 499 & 350 &  264.39 & 630.70 \\ \hline
    $\eta_c(2S)\to h_c(1P)\gamma$ & 111 &  - & 200.63  &  35.19 \\ \hline
  \end{tabular}
\end{table}

After numerically obtaining quarkonia radial wave functions, we then use
\eqref{strongly:hindered:E1:width:formula} and \eqref{New:E1:overlap:integral} to calculate the partial width from ``hard-scattering" mechanism for various $E1$
transition processes.
In Table~\ref{Table:3S1:3PJ:bottomonia}-\ref{Table:E1:charmonia} we enumerate our predictions for various $E1$ transition rates for bottomonia and charmonia family.
For the sake of comparison, we also juxtapose the predictions from the conventional multipole expansion method,
\eqref{conventional:E1:width formula}, together with the measured values whenever available~\cite{Workman:2022ynf}.

For $E1$ transitions in both bottomonium and charmonium, the numerical predictions from the hard scattering mechanism are different from those obtained in the conventional $E1$ formula~\footnote{
It is curious to note that, the predictions from two very different mechanisms appear to differ only by a few times. This is nontrivial since there are no free adjustable parameters
in both \eqref{conventional:E1:width formula} and \eqref{strongly:hindered:E1:width:formula}.}
In the case of the small photon momentum, the multipole expansion picture is expected to hold, and we do not anticipate
that the predictions from our new mechanism are trustworthy anymore. As a matter of fact, in most observed channels with small photon momentum,
the predictions from the conventional formula are much closer to the measured values than those from the ``hard-scattering" mechanism.

We certainly anticipate our predictions become more reliable than the conventional predictions
when the photon momentum exceeds 800 MeV. The transition processes $\chi_{b0,1,2}\to \Upsilon(1S)+\gamma$ are of special interest, since the photon momenta in these processes 
become quite close to  800 MeV. From Table~\ref{Table:3PJ:3S1:bottomonia} we observe that the predictions from the ``hard-scattering" mechanism are considerably larger
than those from the conventional mechanism. Similarly, the momentum of the radiated photon in the process $h_{b}\to \eta_b(1S)+\gamma$ 
is greater than 800 MeV. From Table~\ref{Table:1P1:1S0:bottomonia}, one observes that the prediction from the new mechanism is slightly bigger than that from the conventional one. Hopefully, the future precise measurements of these $E1$ transition channels will unambiguously test whether the ``hard-scattering" mechanism 
is more successful relative to the conventional one or not.

From Table~\ref{Table:3S1:3PJ:bottomonia}-\ref{Table:E1:charmonia}, we also numerically confirm that the predictions from the ``hard-scattering" mechanism 
indeed have a weaker sensitivity to the photon momentum than those from the conventional mechanism. 
This is supported by the fact that the predictions from the new mechanism respects the pattern dictated by the heavy quark spin symmetry to a better extent.

Last but not least, it is worth mentioning a decade-long puzzle, {\it i.e.}, 
the measured $E1$ transition rates for $\Upsilon(3S)\to\chi_{b0,1,2}\gamma$ processes exhibit an anomalous pattern, 
which badly violates the cherished heavy quark spin symmetry.  Clearly this anomaly cannot be accounted by both the
conventional and the new hard-scattering mechanism. There have been some efforts attempting to resolve this puzzle by including
a variety of relativistic corrections in quark potential models~\cite{Moxhay:1983vu,Gupta:1984jb,Grotch:1984gf,Lahde:2002wj,Ebert:2002pp,Badalian:2012mb}.
It is difficult to conceive that the heavy quark spin symmetry can be badly violated for the bottom quark sector, 
especially concerning the fact $v^2\approx 0.1$ for bottomonium.
Obviously, future independent experiments will be essential to finally settle down this puzzle.

\section{Summary}
\label{sec:summary}

The electromagnetic transition between heavy quarkonia is a widely studied topic.
For a very small photon momentum, the dominant $E1$ transition is adequately described by the standard multipole expansion approach.

In sharp contrast to the atomic system, the linear confinement potential in QCD allows the mass difference between the highly excited quarkonium
and the ground state to be as large as 1 GeV. In such situation, the photon can no longer be viewed as ultrasoft, and the
multipole expansion doctrine ceases to work. It is more appropriate to regard the photon momentum as semihard, and adopt 
the hard-scattering mechanism to account for these strongly hindered EM transitions, where a semihard gluon
must be exchanged between heavy quark and antiquark during the single-photon emission.

In this work, we apply the ``hard-scattering'' approach originally developed to tackle the strongly hindered $M1$ transition a decade ago
to deal with the strongly hindered $E1$ transitions between quarkonia.
We derive the compact formulas for the corresponding transition rates in the context of the non-relativistic EFT, 
at the lowest order in the strong coupling and velocity expansion.
Although it is possible to deduce the identical expression from the relativistic QCD, the physics becomes much more transparent from the angle of NREFT.
We have also conducted a detailed numerical study for various $E1$ transition processes between $P$- and $S$-wave bottomonia and charmonia,
and confront our new predictions with the conventional ones. Unfortunately, unlike the $M1$ case, the emitted photon in all the observed $E1$ transition channels in bottomonia and charmonia is not energetic enough to justify the predictions from the hard-scattering mechanism. Nevertheless,  
the photon energy already reaches 800 MeV in the decay channels $\chi_{b0,1,2}\to \Upsilon(1S)+\gamma$ and $h_{b}\to \eta_b(1S)+\gamma$, and the predictions from  
the ``hard-scattering" mechanism are considerably larger those from the conventional approach. Therefore, we hope that
the future precise measurements of these $E1$ transition channels can examine the validity of the ``hard-scattering" mechanism.
In the future, we also plan to further apply this hard-scattering formalism 
to predict the strongly hindered  $D$-to-$P$-wave quarkonia $E1$ transition, and make some testable predictions in close contact with phenomenology.

\begin{acknowledgments}
We are grateful to Jichen Pan for discussions.
This work is supported in part by the National Natural Science Foundation of China under Grants No.~11925506.
The work of  Z. W. M. is also supported in part by the National Science Foundation of China No. 12347145, No. 12022514, No. 12375099 and No. 12047503, and National Key Research and Development Program of China No. 2020YFC2201501 and No. 2021YFA0718304.
The work of J.-Y. Z. is also supported in part by the US Department of Energy (DOE) Contract No. DE-AC05-06OR23177, under which Jefferson Science Associates, LLC operates Jefferson Lab.
\end{acknowledgments}

\appendix
\section{strongly hindered $E1$ transition from QCD}
\label{sec:diagram_approach}

\begin{figure*}[htbp]
\begin{center}
\includegraphics[clip,width=1\textwidth]{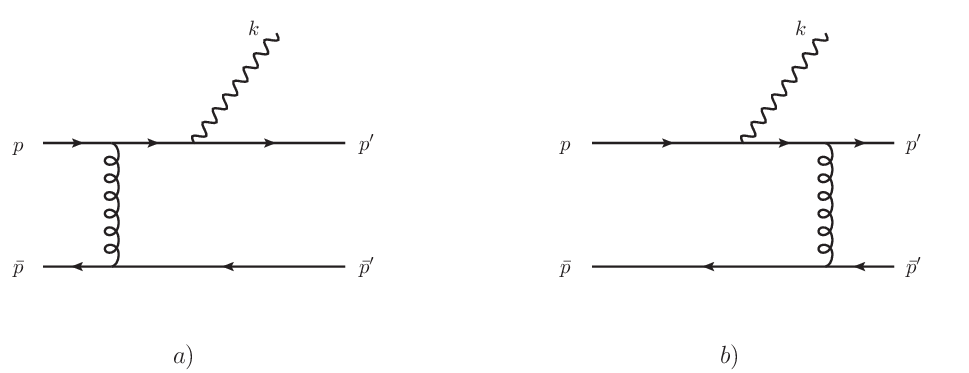}
\caption{The lowest-order Feynman diagrams in QCD+QED that contribute to the strongly hindered $E1$
transition between heavy quarkonia arising from the hard-scattering mechanism. For simplicity, we have suppressed
two mirror diagrams where the photon is emitted from the heavy anti-quark line.}
\label{Hard-scattering:E1:Feynman:diagram:QCD+QED}
\end{center}
\end{figure*}

In this appendix, we show that our key results, \eqref{New:E1:overlap:integral} and \eqref{strongly:hindered:E1:width:formula},
can also be directly reproduced from the underlying relativistic field theory, {\it e.g.}, QCD plus QED.
As a matter of fact, the strongly hindered $M1$ transition in the hard-scattering scattering has been derived from this angle~\cite{Jia:2009yg}.
At the lowest order in $\alpha_s$, there are four Feynman diagrams for $Q\overline{Q}\to Q\overline{Q}+\gamma$ in the hard-scattering mechanism,
two of which are depicted in Fig.~\ref{Hard-scattering:E1:Feynman:diagram:QCD+QED}.
Note the seagull vertex in Fig.~\ref{Hard-scattering:E1:NREFT:diagram} arising from the NREFT
is absent in the underlying relativistic field theory.
The kinematical setting is identical with \eqref{quark:antiquark:momentum:kinematics}.

Let us consider the single-photon transition process $n^{2S+1} P_J\to n'^{2S+1} S_S+\gamma$.
The starting point is \eqref{E1:amplitude:P:wave:to:S:wave}, except the quark amplitude is now computed from underlying
relativistic theory rather than from NREFT:
\bqa
& & \mathcal{A}[n^{2S+1} P_J\to n'^{2S+1} S_S+\gamma  ] =  \int\frac{d^3 {\bf q}}{(2\pi)^3}\int\frac{d^3 {\bf q'}}{(2\pi)^3}
    \frac{1}{N_c} \sum_{i,s_1, s_2, S_z, L_z}
\left\langle {1\over 2} s_1; {1\over 2}s_2 \bigg\vert S\,S_z \right\rangle
\langle S\,S_z; 1\,L_z| J\,J_z \rangle
\nn\\
   &&  \times \sum_{i',s'_1, s'_2}
\left\langle S' \,S'_z   \bigg\vert  {1\over 2} s'_1; {1\over 2}s'_2 \right\rangle
    \widetilde{R}_{nP}(|{\bf q}|)  Y_{1 L_z}(\hat{{\bf q}})  \widetilde{R}^*_{n'S}(|{\bf q'}|)Y^*_{00}(\hat{{\bf q'}})
\label{E1:amplitude:P:wave:to:S:wave:QCD+QED}
\\
&&\times  {\cal A}[Q^i (s_1) \overline{Q}^i(s_2) \to Q^{i'}(s'_1)\overline{Q}^{i'}(s'_2)+\gamma(k)].
\nn
\eqa
The quark amplitude ${\cal A}={\cal A}_a+{\cal A}_b+{\rm two\;mirror\;diagrams}$,
which in Feynman gauge reads
\bseq
\bqa
      {\cal A}_{a}
    &  = &
      ie e_{Q} g_{s}^{2}
            { [\bar{u}(p') \slashed \varepsilon^{*}_{\gamma} (\slashed p'+\slashed k +m) t^a \gamma^{\alpha}u(p)]  \,
            [\bar v(\bar p) t^a \gamma_{\alpha}  v(\bar p')]
            \over
               ((p'+k)^{2}-m^{2} ) ((q- q^\prime -{k\over 2})^2+i\epsilon)
            },
\\
      {\cal A}_{b}
    &  = &
      i e e_{Q} g_{s}^{2}
            { [\bar{u}(p') t^a \gamma^\alpha (\slashed{p}-\slashed k +m) \slashed \varepsilon^{*}_{\gamma} u(p)]\,[\bar{v}(\bar p) t^a\gamma_{\alpha}  v(\bar p')]
            \over
              ((p-k)^2-m^2)  ((q- q^\prime -{k\over 2})^2+i\epsilon)}.
\eqa
\label{QCD:amplitude:Fig:a:Fig:b}
\eseq

It is convenient to employ the covariant projector technique~\cite{Bodwin:2002cfe} to enforce the $Q\overline{Q}$ pair to bear the spin-singlet or spin-triplet quantum numbers.
For the  $Q\overline{Q}$ pair in the initial state, one can make the following substitution in \eqref{QCD:amplitude:Fig:a:Fig:b}:
\bseq
\begin{align}
 \sum_{s_1,s_2} u(p,s_1) \bar v(\bar p,s_2)\langle\frac{1}{2}s_1\frac{1}{2}s_2|00\rangle
  & =
  \frac{-1}{2\sqrt{2}E(E+m)}
  (\slashed{p}+m) {\slashed{P}+2E\over 4E}
  \gamma_{5}
  (\slashed{\bar{p}}-m) \otimes {\boldsymbol{1}_c\over \sqrt{N_c}},
\\
 \sum_{s_1,s_2} u(p,s_1) \bar v(\bar p,s_2)\langle\frac{1}{2}s_1\frac{1}{2}s_2|1S_z\rangle
  & =
\frac{1}{2\sqrt{2}E(E+m)}
   (\slashed{p}+m) {\slashed{P}+2E\over 4E}
 \slashed{\varepsilon}(S_z)
  (\slashed{\bar{p}}-m) \otimes {\boldsymbol{1}_c\over \sqrt{N_c}},
\end{align}
\label{covariant:spin:projector}
\eseq
with $P^2\equiv 4E^2\approx 4m^2$. $\boldsymbol{1}_c$ signifies the $3\times 3$ unit matrix to enforce the projection of the $Q\overline{Q}$ pair
onto the color-neutral state. $\varepsilon(S_z)$ denotes the polarization vector for the unit spin. For the $Q\overline{Q}$ pair in the final state,
one can make the analogous substitution in  \eqref{QCD:amplitude:Fig:a:Fig:b}, by adapting \eqref{covariant:spin:projector} accordingly
with the aid of the identity $v\bar{u}=\gamma^0(u\bar{v})^{\dagger}\gamma^0$.
Note that the Dirac spinors in \eqref{covariant:spin:projector} are normalized
in nonrelativistic convention.

In the hard-scattering picture, both quark and gluon propagators in Fig.~\ref{Hard-scattering:E1:Feynman:diagram:QCD+QED} should be viewed as semi-hard,
with power counting of various momenta expounded in Sec.~\ref{amplitude:quark-level:NREFT}. Similar to the line of reasoning leading to
\eqref{expand:soft:gluon:propagator} and \eqref{expand:soft:quark:propagator},
the quark and gluons propagators in Fig.~\ref{Hard-scattering:E1:Feynman:diagram:QCD+QED} can be simplified as
\bseq
\bqa
  & &\frac{1}{(p'+k)^2-m^2}
   =
   \frac{1}{k\cdot P' + 2k\cdot q'}
   \approx
   \frac{1}{k\cdot P'}\approx \frac{1}{k\cdot P},
\\
 & &\frac{1}{(p-k)^2-m^2}
   =
   -\frac{1}{k\cdot P + 2k\cdot q}
   \approx
  - \frac{1}{k\cdot P},
\\
 & & {1\over (q- q^\prime -{k\over 2})^2+i\epsilon}
  \approx \frac{-1}{
             ({\bf q}'-{\bf q})^2
             +{\bf k}\cdot({\bf q}'-{\bf q})
             -i{\epsilon}}.
\eqa
\label{Expansion:of:propagators:QCD}
\eseq

Implementing the expanded propagators in \eqref{Expansion:of:propagators:QCD},
performing the spin and color projection using \eqref{covariant:spin:projector}, retaining only the linear terms in $\boldsymbol{q}$ and $\boldsymbol{q'}$
in the numerators, the transition amplitude for $n^{2S+1} P_J\to n'^{2S+1} S_S+\gamma$ in \eqref{E1:amplitude:P:wave:to:S:wave:QCD+QED} can be reduced to
\bseq
\bqa
& &  \mathcal{A}[n^{1} P_1(J_z) \to n'^{1} S_0+\gamma  ] =  C_F  {e e_Q  g_s^2\over k\cdot P}
\int\!\!\! \frac{d^3 {\bf q}}{(2\pi)^3}\int\!\!\! \frac{d^3 {\bf q'}}{(2\pi)^3}
\\
&& \times
 \widetilde{R}_{nP}(|{\bf q}|)  Y_{1 J_z}(\hat{{\bf q}})  \widetilde{R}^*_{n'S}(|{\bf q'}|)Y^*_{00}(\hat{{\bf q'}}) \,
\boldsymbol{T}({\bf q'}-{\bf q}) \cdot \boldsymbol{\varepsilon}_{\gamma}^*,
\nn\\
& & \mathcal{A}[n^{3}P_J (J_z) \to n'^{3} S_1(S'_z)+\gamma  ] =
C_F  {e e_Q  g_s^2\over k\cdot P}   \int\!\!\! \frac{d^3 {\bf q}}{(2\pi)^3}\int\!\!\! \frac{d^3 {\bf q'}}{(2\pi)^3}
    \sum_{S_z, L_z}
\langle 1\,S_z; 1\,L_z| J\,J_z \rangle
\\
&& \times
    \widetilde{R}_{nP}(|{\bf q}|)  Y_{1 L_z}(\hat{{\bf q}}) \widetilde{R}^*_{n'S}(|{\bf q'}|)Y^*_{00}(\hat{{\bf q'}})\,
 (\boldsymbol{\varepsilon}^*(S'_z) \cdot \boldsymbol{\varepsilon}(S_z))\,\boldsymbol{T} ({\bf q'}-{\bf q}) \cdot \boldsymbol{\varepsilon}_{\gamma}^*,
\nn
\eqa
\label{Pwave:to:Swave:E1:intermediate:formula:QCD+QED}
\eseq
where the hard-scattering function $\boldsymbol{T} ({\bf q'}-{\bf q})$ has been given in \eqref{kernel:T:definition}.
Hearteningly,  these equations are exactly identical to \eqref{Pwave:to:Swave:E1:intermediate:formula} derived in NREFT.
Following the remaining steps as elaborated in Sec.~\ref{sec:main:body},  we then reproduce the key formula of this work,
 \eqref{New:E1:overlap:integral} and \eqref{strongly:hindered:E1:width:formula}.


\end{document}